\documentclass[aps,prl,twocolumn,groupedaddress]{revtex4}

\usepackage{graphicx}
\usepackage{dcolumn}
\usepackage{bm}

\begin{document}

\title{\large Signature of a crossed Andreev reflection effect (CARE)\\
in the magnetic response of $YBa_2Cu_3O_{7-\delta}$ junctions\\
with the itinerant ferromagnet $SrRuO_3$}

\author{P. Aronov and G. Koren }
\affiliation{Physics Department, Technion - Israel Institute of
Technology Haifa, 32000, ISRAEL}

\email{gkoren@physics.technion.ac.il}
\homepage{http://physics.technion.ac.il/~gkoren}

\date{\today}
\def\bfig {\begin{figure}[tbhp] \centering}
\def\efig {\end{figure}}

\normalsize \baselineskip=4mm  \vspace{15mm}

\begin{abstract}
Magnetic properties of SFS and SF ramp-type junctions with
$YBa_2Cu_3O_{7-\delta}$ (YBCO) electrodes (S), and the itinerant
ferromagnet $SrRuO_3$ (SRO - F), were investigated. We looked for
a crossed Andreev reflection effect (CARE) in which an electron
from one magnetic domain in F is Andreev reflected as a hole into
an adjacent, oppositely polarized, domain while a pair is
transmitted into S. CARE is possible in SRO since the width of its
domain walls is of the order of the YBCO coherence length (2-3
nm). Our junctions behave as typical magnetic tunneling junctions,
as the conductance spectra were always asymmetric, and a few
showed bound state peaks at finite bias that shifted with field
according to the classical Tedrow and Meservey theory. In many of
our SFS junctions with a barrier thickness of 10-20 nm, a
prominent zero bias conductance peak (ZBCP) has been observed.
This peak was found to decrease \textit{linearly} with magnetic
field, as expected for Andreev and CARE scattering. In contrast,
in SF junctions, the observed ZBCP was found to decrease versus
field almost \textit{exponentially}, in agreement with the
Anderson-Appelbaum theory of scattering by magnetic states in F.
Thus, transport in our SFS and SF junctions depends strongly on
the size of the F layer. We also found that in both cases, the
ZBCP height at zero field decreased with increasing magnetic order
of the domains in F, in agreement with the CARE mechanism.
\end{abstract}


\maketitle

Properties of SFS ramp type junctions of $YBa_2Cu_3O_{7-\delta}$
(YBCO) electrodes (S) and $SrRuO_3$ (SRO) barrier (F)  have been
investigated by two groups more than ten years ago
\cite{Char-SRO,Juelich}. Both groups have found that the normal
resistance values of their junctions show two distinct phenomena.
One, that the observed values had a large spread from a few tens
of Ohms to a few hundred Ohms, and the other that they were two or
more orders of magnitude higher than the expected Ohmic resistance
of the SRO film. The high normal resistance was therefore assumed
to originate at the YBCO/SRO interfaces, and more specifically
attributed to oxygen disorder and depletion near the interfaces
\cite{Juelich,Char-interface,Olsson}. D\"{o}mel \textit{et al.}
attributed the high normal resistance to an insulating interface
layer through which quasiparticle tunnelling via localized states
occurred \cite{Juelich}. A second explanation to the high normal
resistance was given by Antognazza \textit{et al.} who attributed
it to interface stress created by thermal expansion mismatch in
the junctions \cite{Char-SRO}. Conductance spectra of YBCO based
SFS junctions with SRO and $CaRuO_3$ (CRO) barriers were also
measured by Antognazza \textit{et al.} \cite{Char-conductance}.
With both type of barriers they found a zero bias conductance peak
(ZBCP) in the center of a tunneling-like gap structure. The
critical current density however, in the junctions with the CRO
barrier persisted up to a barrier thickness of 50\,nm, while that
of the junctions with the SRO barrier vanished abruptly already at
a thickness of 25\,nm \cite{Char-CRO,Char-interface}. Since SRO is
an itinerant ferromagnet below $\sim 150$K \cite{Klein,Zakharov}
while CRO is a paramagnet, the different behavior of the critical
current density indicates that the magnetic properties of the
barrier layer play an important role in the transport of these
junctions. \\

In the present study we revisit the same type of junctions of YBCO
and SRO, with a special focus on transport properties which are
affected by the magnetic nature of the barrier material. In the
absence of a critical current, transport in junctions at voltage
bias values below the energy gap of the superconductor is
controlled by Andreev scattering. When the barrier material is
fully spin polarized in one direction, no Andreev transport is
possible. If however, the ferromagnetic barrier has many domains
with opposite polarizations, a crossed Andreev reflection effect
(CARE) is possible \cite{Deutscher}. This effect can occur at the
intersection of the domain walls and the YBCO electrodes at the
interfaces, provided the value of the domain walls width is
similar to that of the superconductor's coherence length $\xi$
(2-3\,nm for optimally doped YBCO). We have chosen to study
junctions with an SRO barrier since the domain wall width of this
highly anisotropic ferromagnet is very narrow ($\sim$3 nm only
\cite{TEM,Michaelklein}), and fulfils the above condition. As will
be described in the following, our conductance spectra results
under magnetic fields provide supportive evidence for the
existence of CARE in our junctions. It should be noted that
recently CARE was observed by Beckmann \textit{et al.} in
conventional FSF junctions made of two closely spaced Fe nanowires
in contact with an Al electrode \cite{Beckmann}. They found that
the resistance difference between parallel and antiparallel
magnetization of the Fe electrodes when the Al electrode was in
the superconducting state, decay with increasing distance of the
Fe electrodes up to about twice the coherence length of Al, in
agreement with the CARE phenomena. Technically however, it is
impossible to reproduced this kind of study in the high
temperature superconductors due to their extremely short coherence
length.\\

We prepared the YBCO based ramp-type junctions with SRO on (100)
$SrTiO_3$ (STO) wafers with a ramp angle of $\sim 35^\circ$. This
was done by a multi-step process, where the epitaxial thin film
layers are prepared by laser ablation deposition, patterning is
done by deep UV photolithography, and etching by Ar ion milling
\cite{Nesher}. The YBCO films had \textit{c-axis} orientation
normal to the wafer. The thickness of the base and cover
electrodes was kept constant at 80\,nm, while the SRO thickness on
different wafers ranged between 4\,and 80\,nm. On each wafer we
patterned ten identical junctions along the (100) direction, with
a width of $5\,\mu m$. Finally, a gold layer was deposited and
patterned to produce the $4\times 10$ contact pads for the 4-probe
transport measurements. The YBCO electrodes of our junctions had
oxygen content close to optimal doping  with a $T_c$ of 88-89\,K.
The quality of our junctions fabrication process was tested by
measuring the critical current density $J_c$ of "shorts"
(junctions without any barrier). We found
$J_c(77K)\sim1\times10^6\,A/cm^2$ which is reasonable compared to
$J_c(77K)\sim3-5\times10^6\,A/cm^2$ found in the best blanket
films.\\

\begin{figure}
\includegraphics[height=6.5cm,width=9cm]{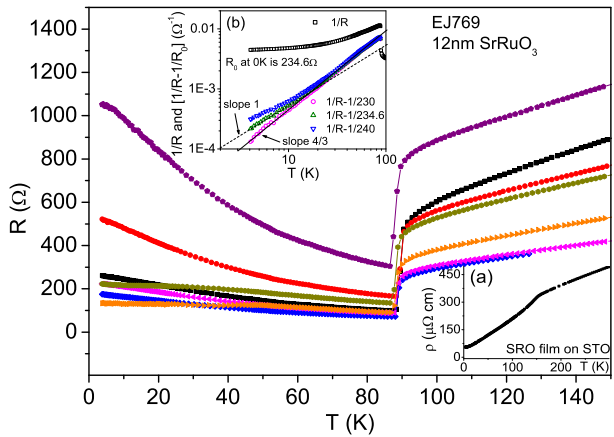}
\caption{\label{fig:epsart}(Color online) Resistance versus
temperature of several junctions on a single wafer. Inset (a):
resistivity versus temperature of a 90 nm thick $SrRuO_3$ film on
(100) STO. Inset (b): inverse resistivity of one of the low
resistance junctions (J6) versus T on a log-log scale (solid line
- $T^{4/3}$, dashed line - $T^1$).}
\end{figure}

Fig.~1 shows the resistance versus temperature of several junctions
on a single wafer. The spread of resistance values at temperatures
above $T_c$ is \textit{extrinsic} and due to the different length of
the YBCO leads to the junctions. At low temperatures, the resistance
values have an \textit{intrinsic} large spread of about one order of
magnitude. These resistance values are also several order of
magnitude higher than the calculated $10\,m\Omega$ Ohmic resistance
of the junction obtained by using the resistivity of the $SrRuO_3$
film (see inset (a) to Fig.~1). These observations are similar to
the results reported previously on the same kind of junctions by
other groups as discussed in the introduction to this paper
\cite{Char-SRO,Juelich}. Unlike previous results however, our
junctions generally had a critical current up to a barrier
thicknesses of $\sim10\,$nm, but were resistive at higher barrier
thicknesses. Antognazza \textit{et~al.} found critical currents with
a barrier thickness of up to 20\,nm \cite{Char-SRO}. This is
possibly due to microshorts or tunneling via oxygen disorder states
in their junctions. D\"{o}mel \textit{et al.} have found that the
inverse resistance difference $1/R-1/R_0$ of their junctions (where
$R_0$ is the extrapolated resistance to $T=0$), varies versus
temperature as $T^{4/3}$. They concluded that this indicates
tunneling via one and two localized states in the barrier
\cite{Juelich,Glazman}. We basically observed a similar behavior as
shown in inset (b) to Fig.~1. As one can see however, the results
are very sensitive to $R_0$ and the $T^{4/3}$ behavior is not
obtained with the extrapolated $R_0$, but a value close to it.
Furthermore, in D\"{o}mel \textit{et al.} study, there is no data
between 5-20\,K. If we use our data in this temperature range, and
with the extrapolated $R_0$ value, we find a \textit{linear}
dependence versus T. Thus we believe that tunneling via two
localized states is \textit{not} the dominant transport mechanism
in our junctions.\\

\begin{figure}
\includegraphics[height=6.5cm,width=9cm]{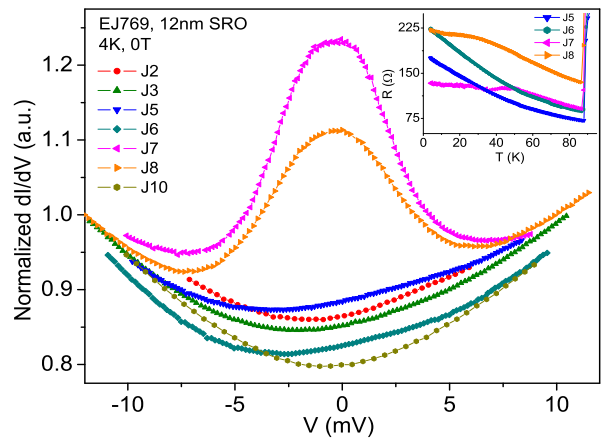}
\caption{\label{fig:epsart}(Color online) Normalized conductance
spectra of several junctions on a single wafer. Inset: zoom up on
the resistance versus temperature of a few junctions. }
\end{figure}

\begin{figure}
\includegraphics[height=6.5cm,width=9cm]{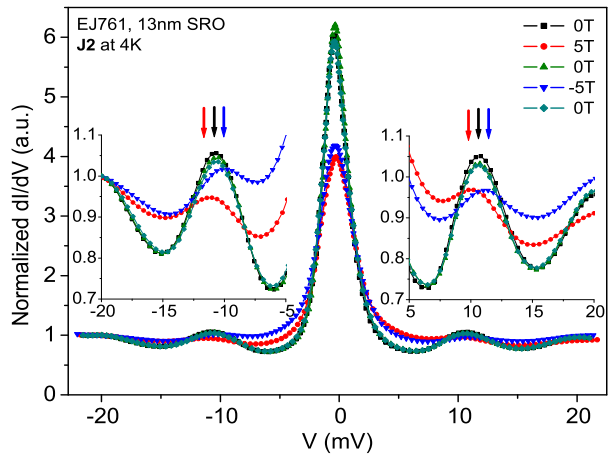}
\caption{\label{fig:epsart}(Color online) Normalized conductance
spectra of a junction under zero field cooling (ZFC) to 4K
followed by field cycling to 5T, 0T, -5T and back to 0T. Insets:
zoom up on the spectra of the bound states near 10 and -10 mV. }
\end{figure}

Fig.~2 shows the normalized conductance spectra at low temperature
of several junctions on a single wafer with a barrier thickness of
12 nm. One observes that two junctions have a ZBCP inside a
tunneling-like structure, while the others have only the
tunneling-like behavior. The ZBCP can be attributed to either
Andreev reflections or scattering by magnetic states
\cite{Andreev,Anderson,Appelbaum}. Since our junctions are
orientated along the \textit{a} or \textit{b} axes of the YBCO
electrodes, the observed ZBCP is \textit{not} due to the well known
bound states which are formed along the node direction in a d-wave
superconductor. In addition, we point out that there is a
correlation between the appearance of a ZBCP and the behavior of the
resistance curves below $T_c$ (see the inset to Fig.~2). For
junction J7 and J8, where a ZBCP was observed, the R versus T curves
show a change of slope (a cusp) and not a monotonic increase with
decreasing temperature as for instance is found in junctions J5 and
J6. Another distinct feature in Fig.~2 is that the spectra are
clearly asymmetric. This asymmetry is typical of ferromagnetic
tunneling junctions due to the opposite shifts of the spectra for up
and down spins, and the non zero spin polarization of the magnetic
electrode \cite{Tedrow-Meservey}. Further support to the fact that
our junctions behave as classical magnetic tunneling junctions, is
found in Fig.~3. In this figure one sees the prominent ZBCP and its
suppression under applied magnetic fields. But first we shall focus
on the bound state peaks observed at about $\pm$10\,mV. As can be
seen by the zoom up on these parts of the spectra, a positive
magnetic field shifts the peak to negative bias, and vice versa. The
total measured shift for fields of $\pm$5 T is $\sim$1.2\,mV. The
expected shift for an SFS junction is $4\mu H$ (1.28 mV here)
\cite{Tedrow-Meservey}. Thus for a bound state of energy $\Delta_1$,
the expected shift $4\mu H/2\Delta_1$ should be equal to the
measured shift $\sim1.2/\Delta$ where $\Delta$ is the gap energy.
Therefore, $2\Delta_1\sim \Delta$ and if $\Delta$ of YBCO is
$\sim$20 meV then $\Delta_1\sim$10 meV, in agreement with
the peaks bias of the bound state in Fig.~2.\\

\begin{figure}
\includegraphics[height=6.5cm,width=9cm]{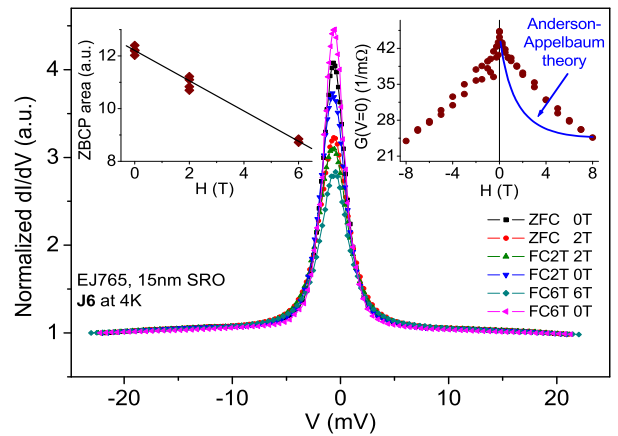}
\caption{\label{fig:epsart}(Color online) Normalized conductance
spectra under various fields and field cooling conditions to 4K.
Insets: the conductance peak area and the conductance at zero bias
versus field. The solid curve is a typical theoretical prediction
of the Anderson-Appelbaum model.}
\end{figure}

Fig.~4 shows a few conductance spectra under different fields and
field cooling conditions. There is almost no effect on the spectra
at any given field larger than about 0.1T, whether it was obtained
under zero field cooling (ZFC) or field cooling (FC). The insets
of Fig.~4 show the ZBCP area above the background conductance, and
the conductance at zero bias ($G_0\equiv G(V=0)=1/R(V=0)$) versus
field. Surprisingly, both features show a \textit{linear} decrease
with increasing field, except maybe for fields near zero field.
The expected decrease of $G_0$ versus field due to scattering by
magnetic states in junctions was calculated by Appelbaum and found
to be almost exponential \cite{Appelbaum67}. A similar behavior,
but with a more gradual decrease versus field, was found also in
experiments done in Ta-I-Al tunnel junctions
\cite{Appelbaum-Shen}. Thus the linear $G_0$ versus $H$ result in
our junctions points to a different scattering mechanism. A
theoretical calculation of the current and magnetoresistance in
FSF junctions due to CARE was recently published, but it did not
include conductance spectra which are relevant in the present
study \cite{Yamashita}. The closest theoretical calculation we
could find for a ZBCP behavior versus $H$ was in a study by Tanaka
\textit{et al.} \cite{Tanaka}. They calculated the conductance
spectra for the node direction in the cuprates using the extended
BTK model for the d-wave superconductors. Clearly, the resulting
ZBCP is due to bound states because of the sign change of the
order parameter, and not to the CARE process. Nevertheless, the
basic scattering mechanism is still Andreev reflections, and
therefore a comparison of our data with their results is
justified. Extracting $G_0$ from their conductance spectra at
different fields, one finds a clear linear decrease with field.
Hence, our data is consistent with this behavior, and it is likely
that simple Andreev and CARE play a dominant role in the transport
of our SFS junctions.\\

Fig.~5 shows a series of conductance spectra in another junction.
These spectra were obtained under various fields starting with ZFC
to 4\,K, ramping up to 8\,T, and going back to zero field. Here
again as in Fig.~4, one finds a linear decrease of the ZBCP height
versus field as shown in the inset to this figure, but the ZFC data
point seems to stand out. Clearly the measured ZBCP height after ZFC
is larger than that obtained after field cycling to 8\,T and back to
0\,T. To understand this behavior, we note that during the ZFC
process, the SRO barrier layer becomes ferromagnetic with many
domains and domain walls as shown schematically in the upper-right
corner of the inset. As a result, the contribution of CARE to the
conductance which depends on the number of domain wall intersections
with the S electrodes, should be higher than the conductance after
field cycling. This is so since the magnetic memory after the field
cycling reduces the number of domains as shown schematically in the
lower-left corner of the inset to Fig.~5. Because this is exactly
the observed result, we conclude that CARE is responsible for the
excess conductance at zero field under ZFC as shown in the inset to
Fig.~5. We stress that this phenomenon is a unique signature of
CARE, which can\textit{ not}
be explained by the standard Andreev reflections.\\

\begin{figure}
\includegraphics[height=6.5cm,width=9cm]{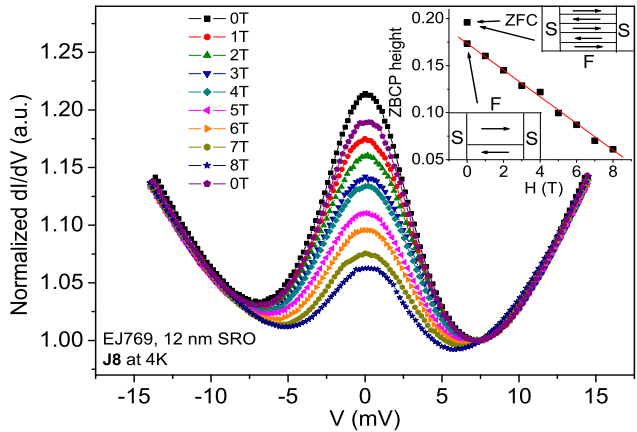}
\caption{\label{fig:epsart}(Color online) Normalized conductance
spectra under ZFC to 4 K, followed by field ramping to 8 T and
back to 0 T. Inset: the ZBCP height above the background
conductance $G_0-G_B$ versus field. $G_B$ was taken as the
conductance value at V=0 of the straight line connecting the two
minima of each spectrum.  }
\end{figure}

Next, we decided to look at the limit of a very thick barrier. Since
the Ohmic resistivity of SRO is quite small at 4\,K, only $\sim
50\,\mu\Omega cm$ as shown in inset (a) of Fig.~1, we chose to study
SF rather than SFS junctions. In this case, the F electrode
"thickness" or size is almost infinite, and therefore its relatively
weak itinerant ferromagnetism should be enhanced \cite{Klein}. The
resulting conductance spectra of a typical SF junction are shown in
Fig.~6. The spectra and the ZBCP height values versus field were
obtained by ZFC to 4\,K, ramping to 6\,T, and going back to 0\,T.
The ZBCP inside a gap-like structure is still present, but its
magnitude is greatly reduced compared to the previous data in SFS
junctions. The interesting feature here is that already at 4\,T, the
ZBCP is almost fully suppressed. Even more amazing is the field
dependence of the ZBCP height as shown in the inset to Fig.~6. This
is clearly \textit{nonlinear}, and rather close to exponential
decay. Actually, this decay is very similar to that predicted by the
Anderson-Appelbaum (AA) theory of scattering by magnetic states
close to the interface with a superconductor
\cite{Anderson,Appelbaum,Appelbaum-Shen}. It should be noted that in
the AA model this decay is due to the increased Zeeman splitting of
the ZBCP \cite{Appelbaum67}. We however, have never observed
splitting of the ZBCP, and this could be due to a larger magnetic
relaxation rate in SRO which broadens this peak and smears the
splitting. It is therefore concluded that the almost exponential
decay versus field indicates that the dominant transport mechanism
now is not Andreev scattering, but rather magnetic scattering. We
conclude that the size of the F electrode plays an important role in
determining the transport properties of our junctions. When the F
electrode is thin as in the previous results of SFS junctions
(10-20\,nm), its ferromagnetism is weak and the proximity effect by
the S electrodes makes it even weaker. The opposite is true when the
F electrode size is large as in the SF junctions case. Then the
proximity penetration of superconductivity into F is small compared
to the mean free path in the F electrode (which is unlimited now by
the junction length), the ferromagnetic order in F is robust, and
the transport in the junction is controlled
mostly by magnetic scattering.\\

\begin{figure}
\includegraphics[height=6.5cm,width=9cm]{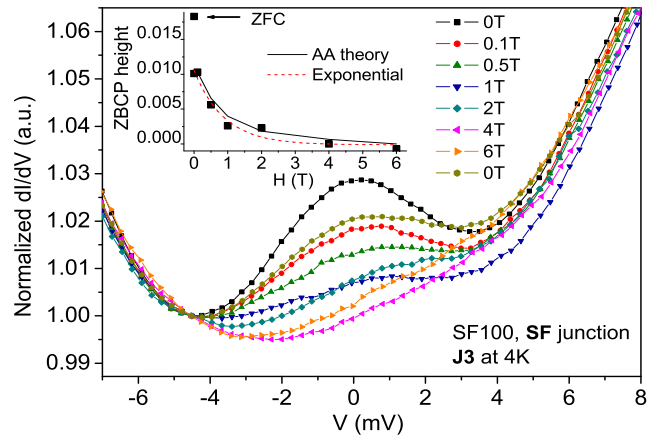}
\caption{\label{fig:epsart} (Color online) Normalized conductance
spectra of an SF junction under ZFC to 4K, and various magnetic
fields at low temperature. Inset: the ZBCP peak height $G_0-G_B$
versus field. The curves are a typical Anderson-Appelbaum fit and
an exponential fit. }
\end{figure}

We note that the value of the ZBCP height after ZFC from room
temperature to 4\,K is still much larger than its value after field
cycling to 6\,T and back to 0\,T, similar to the result in the SFS
junction of Fig.~5. It is tempting to attribute this behavior to
CARE as before, but then the absence of a linear decreasing
component of the ZBCP height versus field which originates in
Andreev scattering, will have to be explained. According to Yokoyama
\textit{et al.} who calculated the conductance spectra due to
magnetic scattering in SN junctions with  a d-wave superconductor
\cite{Tanaka3}, an enhanced magnetic scattering rate (by the higher
magnetic disorder after ZFC in the present study) would decrease
rather than increase the ZBCP. Since this is opposite to
observation, it seems that we are still dealing with suppression of
the ZBCP height due to CARE here (from the magnetically disordered
ZFC state to the more ordered state after field cycling, similar to
the result in  Fig.~5). Apparently, in the SF case where the
ferromagnetism of F is robust, the linear suppression of the ZBCP
height versus field is much enhanced and terminates at a much
smaller applied field. This leaves only the exponential decay versus
field due to magnetic scattering as the
dominant process.\\

In conclusion, we have found significant magnetic effects in the
transport properties of SFS and SF junctions of YBCO and the SRO
ferromagnet. \textit{i}) We observed an asymmetry in the
conductance spectra, and shifts of bound state peaks with field,
which are typical of magnetic tunneling junctions. \textit{ii}) In
both type of junctions a prominent ZBCP was observed. Its height
decreased linearly with increasing field in SFS junctions, but
almost exponentially in the SF case. The ZBCP height dependence on
$H$ originated in normal Andreev and CARE in the SFS junctions,
but was dominated by magnetic scattering in the SF junctions.
\textit{iii}) The observation of a higher ZBCP height at 0\,T
after ZFC as compared to the value after field cycling is due to
the higher magnetic disorder after ZFC in both SFS and SF
junctions. This is a strong signature of a CARE phenomenon in our
junctions. Finally, we note that a calculation of the conductance
spectra under fields in SF and SFS junctions is needed for a more
quantitative comparison with the present results.\\

{\em Acknowledgments:}  The authors are grateful to G. Deutscher,
L. Klein, O. Millo and E. Polturak for useful discussions. This
research was supported in part by the Israel Science Foundation
(grant \# 1564/04), the Heinrich Hertz Minerva Center for HTSC,
the Karl Stoll Chair in advanced materials, and by the Fund for
the Promotion of Research at the Technion.\\

\bibliography{AndDepBib.bib}

\bibliography{apssamp}

\end{document}